\documentclass[prl,twocolumn,showpacs,superscriptaddress,amsfonts,amsmath,floatfix]{revtex4}
\usepackage{graphicx}
\newcommand{\Journal}[4]{#1 {\bf #2}, #3 (#4)}
\newcommand{\PRL}{Phys. Rev. Lett.}

\newcommand{\Science}{Science}
\newcommand{\PLA}{Phys. Lett. A}
\usepackage[normalem]{ulem}
\begin{document}
\title{Wave functions of the super Tonks-Girardeau gas and the trapped 1D hard sphere Bose gas}
\author{M. D. Girardeau}
\email{girardeau@optics.arizona.edu}
\affiliation{College of Optical Sciences, University of Arizona, Tucson, AZ 85721, USA}
\author{G.E. Astrakharchik}
\email{astrakharchik@mail.ru}
\affiliation{Departament de F\'{\i}sica i Enginyeria Nuclear, Campus Nord B4, Universitat Polit\`ecnica de Catalunya, E-08034 Barcelona, Spain}
\date{\today}
\begin{abstract}
Recent theoretical and experimental results demonstrate a close connection between the super Tonks-Girardeau (sTG) gas and a 1D hard sphere Bose (HSB) gas with hard sphere diameter nearly equal to the 1D scattering length $a_{\text{1D}}$ of the sTG gas, a highly excited gas-like state with nodes only at interparticle separations $|x_{j\ell}|=x_{\text{node}}\approx a_{\text{1D}}$. It is shown herein that when the coupling constant $g_B$ in the Lieb-Liniger interaction $g_B\delta(x_{j\ell})$ is negative and $|x_{12}|\ge x_{\text{node}}$, the sTG and HSB wave functions 
for $N=2$ particles 
are not merely similar, but identical; the only difference between the sTG and HSB wave functions is that the sTG wave function allows a small penetration into the region $|x_{12}|<x_{\text{node}}$, whereas for a HSB gas with hard sphere diameter $a_{\text{h.s.}}=x_{\text{node}}$, the HSB wave function vanishes when $|x_{12}|<a_{\text{h.s.}}$. 
Arguments are given suggesting that the same theorem holds also for $N>2$.  
The sTG and and HSB wave functions for $N=2$ are given exactly in terms of a parabolic cylinder function, and for $N\ge 2$, $x_{\text{node}}$ is given accurately by a simple parabola.  
The metastability of the sTG phase generated by a sudden change of the coupling constant from large positive to large negative values is explained in terms of the very small overlap between the ground state of the Tonks-Girardeau gas and collapsed cluster states.
\end{abstract}
\maketitle
If an ultracold atomic vapor is confined in a de Broglie wave guide with
transverse trapping so tight and temperature so low that the transverse
vibrational excitation quantum is larger than available
longitudinal zero point and thermal energies, the effective dynamics becomes
one-dimensional (1D) \cite{Ols98,PetShlWal00}. 3D Feshbach resonances \cite{Rob01}
allow tuning to the neighborhood of
1D confinement-induced resonances (CIRs) \cite{Ols98,BerMooOls03}, where
the 1D interaction is very
strong, leading to strong short-range correlations, breakdown of
effective-field theories, and emergence of highly-correlated $N$-body
ground states. In the bosonic case with zero-range repulsion $g_B\delta(x_j-x_\ell)$
with coupling constant $g_B\to +\infty$, the
Tonks-Girardeau (TG) gas, the exact
$N$-body ground state was determined in 1960 by a
Fermi-Bose (FB) mapping to an ideal Fermi gas \cite{Gir60}, leading to
``fermionization'' of many properties of this Bose system, as recently
confirmed experimentally \cite{Par04,Kin04}. It is now known
\cite{CheShi98,GirOls03,GirNguOls04} that the FB mapping
is of much greater generality; when supplemented by an inversion and sign
change of the coupling constant, it provides a mapping between
the $N$-body energy eigenstates of a 1D Bose gas with delta-function
interactions $g_{B}\delta(x_j-x_\ell)$ of any strength [Lieb-Liniger (LL) gas \cite{LieLin63}] and those
of a spin-aligned Fermi gas.

Practically all experiments on ultracold Bose gases with tight transverse trapping (1D regime) are for the case where, in addition to tight transverse trapping, there is weak longitudinal trapping by a harmonic oscillator potential $m\omega^2 x^2/2$ where $x$ is the longitudinal coordinate. In that case the strength of effective 1D LL boson-boson interactions is characterized by a dimensionless coupling constant $\lambda=g_B/2^{3/2}\hbar\omega x_{\text{osc}}$ where $x_{\text{osc}}=\sqrt{\hbar/m\omega}$ is the oscillator length. 
For $\lambda\to +\infty$ the exact ground state is the trapped TG gas, and for $\lambda\to -\infty$ the same TG state
is still an exact eigenstate (though now highly excited), since 
both cases map to the ideal Fermi gas \cite{Gir60,CheShi98}. 
If $\lambda$ is negative and finite, the ground state is very different; it is McGuire's collapsed cluster state (bright soliton) \cite{McG64}. However, if $\lambda$ is very large and negative, it was first shown theoretically \cite{AstBluGioGra04,BatBorGuaOel05,AstBorCasGio05} that the system is metastable against collapse to McGuire's cluster state in spite of the very strong attractive interactions, and exhibits a strong similarity with a trapped 1D hard sphere Bose (HSB) gas with hard sphere diameter $a_{\text{h.s.}}>0$ nearly equal to the 1D scattering length $a_{\text{1D}}$ of the system with $\lambda<0$. It is even more highly correlated than the TG gas, and hence was named \cite{AstBluGioGra04} the ``super Tonks-Girardeau'' (sTG) gas. By changing $\lambda$ \emph{suddenly} from large positive to large negative values by passing through the CIR, the sTG gas was recently created experimentally and
shown to have properties in agreement with the theoretical predictions \cite{Haletal09}.

{\it Exact solution for} $N=2$: In the simplest case of two bosons with LL interaction $g_{B}\delta(x_1-x_2)$ in a harmonic trap potential $m\omega^{2}(x_1^{2}+x_2^{2})/2$, all energy eigenstates and eigenvalues can be easily obtained. The wave function for the center of mass (c.m.) coordinate $X=(x_1+x_2)/2$ is  $\psi_{\text{c.m.}}=\exp[-X^2/x_{\text{osc}}^2]$ and its energy is $E_{\text{c.m.}}=\hbar\omega/2$ assuming that the c.m. mode is unexcited. Denote the relative wave function by $\phi(x)$ with $x=x_{1}-x_{2}$.
The excited eigenstates $\psi_\nu(x_1,x_2)$ are most easily found by FB mapping \cite{Gir60}. The Fermi-Bose (FB) mapping relation \cite{Gir60} reduces to $\phi_{B}(x)=-\mbox{sgn}(x)\phi_{F}(x)$. $\phi_B$ satisfies the same harmonic oscillator (HO) wave equation as $\phi_F$ for $x\ne 0$, and at $x=0$ both satisfy a derivative condition which for $\phi_F$ reads
\begin{equation}\label{jump}
{\phi_{F}}^{'}(0\pm)=(\mu g_B/\hbar^2)\phi_{F}(0+)=-(\mu g_B/\hbar^2)\phi_{F}(0-)
\end{equation}
where $\mu$ is the effective mass $m/2$. Since $\phi_F$ is fermionic, it is
an odd function, but nevertheless it does not vanish at $x=0$;
instead, it changes sign there, but its derivative is continuous.
For $x\ne 0$, $\phi_F$ satisfies the free particle HO
wave equation valid for $x\ne 0$ supplemented by the cusp boundary
condition (\ref{jump}).
Two different solutions of the HO wave equation having nonzero value and
nonzero derivative at $x=0$ are required, such that the boundary
conditions of (\ref{jump}) are satisfied.
These cannot be satisfied by the usual Hermite-Gaussian solutions;
instead, solutions which are essentially analytic continuations of
these to nonintegral quantum number $n$ are required. Since $\phi_F$
is necessarily an odd function of $x$, it is sufficient to obtain a
solution $y$ of the differential equation only for $x>0$,
taking $\phi_{F}(x)=y(q)$ for $x>0$ and $\phi_{F}(x)=-y(|q|)$
for $x<0$, where $x=qx_{\text{osc}}$. The necessary solution
vanishing and integrable as $q\to +\infty$ is a parabolic cylinder function
$D_{\nu}(q)$ expressible in terms of confluent hypergeometric series $\Phi(\alpha,\gamma;z)$
which is a sum of two terms, one even and the other odd in $q$ \cite{Gra80,MorFes53,Cir01,Fra03,TemSolSch08}.
\begin{eqnarray}\label{parabolic}
&&D_{\nu}(q)=2^{\frac{\nu}{2}}e^{-\frac{q^2}{4}}\left[\frac{\sqrt{\pi}}{\Gamma(\frac{1}{2}
-\frac{1}{2}\nu)}\Phi\left(-\frac{\nu}{2},\frac{1}{2};\frac{q^2}{2}\right)\right.\nonumber\\
&&-\left.\frac{q\sqrt{2\pi}}{\Gamma(-\frac{\nu}{2})}
\Phi\left(\frac{1}{2}-\frac{1}{2}\nu,\frac{3}{2};\frac{q^2}{2}\right)\right]\ .
\end{eqnarray}
This solution vanishes like $q^{\nu}e^{-q^{2}/4}$ as $q\to +\infty$; it diverges like
$|q|^{-\nu-1}e^{q^{2}/4}$ as $q\to -\infty$, but that is of no
consequence since we require it only for $q>0$. The relative energy
eigenvalues are $E_{\text{rel}}=(\nu +\frac{1}{2})\hbar\omega$ where the
nonintegral quantum numbers $\nu$
are determined by the derivative condition (\ref{jump}). This leads to the following
transcendental equation for the allowed values \cite{Fra03}:
$\Gamma(\frac{1}{2}-\frac{1}{2}\nu)/\Gamma(-\frac{1}{2}\nu)=-\lambda$
in terms of the previously defined dimensionless coupling constant $\lambda$.
Multiplying by the signum function so as to obtain the bosonic (even)
solution $\phi_{B}(x)=-\text{sgn}(x)\phi_{F}(x)$ one obtains a bosonic
solution continuous and nonzero at $x=0$ but with a cusp (derivative
sign change) there.
The total wave function including the c.m. factor can be written as
$\psi_{B\nu}(x_1,x_2)=\exp[-(x_1^2+x_2^2)/2x_{\text{osc}}^2]f_\nu(|q|)$ where
$f_\nu(|q|)=D_\nu(|q|)e^{q^{2}/4}$ and $q=(x_1-x_2)/x_{\text{osc}}$. The exponential growth of $e^{q^{2}/4}$ as $|q|\to\infty$
is cancelled by the exponential decrease of $D_\nu(|q|)$ like $e^{-q^{2}/4}$.

The first excited state is the one with $\nu\to 1$ as $|\lambda|\to\infty$ and has only one node at $|x_1-x_2|=x_{\text{node}}$.
For $|\lambda|=\infty$, $\phi_F$ is just the first excited HO state $qe^{-q^2/2}$ \cite{GirWriTri01} with a node only at the origin, and for finite negative $\lambda$, $x_{\text{node}}>0$.
It is clear from Figs. 5(b,c,d) of \cite{Fra03} that $\phi_{B}$ is almost linear in the region $0\le |x|\le x_{\text{node}}$
for all negative $\lambda$, so it is well approximated there by a parabola, which can by found explicitly from the leading terms
of the hypergeometric series in Eq. (\ref{parabolic}) and the Gaussian prefactor.
By Eq. (\ref{parabolic}),
\cite{Gra80}, and the transcendental equation for $\nu$ one finds
$\phi_{B}\approx c\{|q|-\frac{a_{\text{1D}}}{x_{\text{osc}}}[1-(\frac{\nu}{2}+\frac{1}{4})q^2)+\cdots]\}$ where $c$ is a normalization
constant. The node is then at
$x_{\text{node}}=a_{\text{1D}}[1-(\frac{\nu}{2}+\frac{1}{4})(\frac{a_{\text{1D}}}{x_{\text{osc}}})^2+\cdots]$, which is very close
to $a_{\text{1D}}$ when $a_{\text{1D}}\ll x_{\text{osc}}$, the case when $|\lambda|=x_{\text{osc}}/\sqrt{2}a_{\text{1D}}\gg 1$.

The ground state is very different from the excited states, being a collapsed state which is an analog, for the trapped system, of
McGuire's cluster state \cite{McG64}. It is an even solution also expressible
in terms of a $D_\nu$, but one whose energy approaches $-\infty$ as $g_B\to -\infty$ ($a_{\text{1D}}\to 0+$); see Fig. 5 of \cite{Fra03}.
For $a_{\text{1D}}\to 0+$ it is well approximated by
$\psi_{B0}\approx\exp(-|x_1-x_2|/a_{\text{1D}})\exp[-(x_1^2+x_2^2)/2x_{\text{osc}}^2]$; in fact,
this expression satisfies the $x_{12}\to 0$ contact condition exactly and
becomes exact for all $(x_1,x_2)$ in both limits $a_{\text{1D}}\to 0+$ (total collapse) and $a_{\text{1D}}\to +\infty$ (ideal Bose gas).

{\it Comparison with trapped hard sphere gas:} The
recent theoretical \cite{AstBluGioGra04,BatBorGuaOel05,AstBorCasGio05} and experimental \cite{Haletal09} results
demonstrate a close connection between the sTG gas and a
1D hard sphere Bose (HSB) gas with hard sphere diameter $a_{\text{h.s.}}$ nearly equal to $a_{\text{1D}}$, the 1D scattering length of the
sTG gas. Actually, when 
all interparticle separations $|x|$ are larger than the nodal position $x_{\text{node}}$, the sTG and
HSB wave functions are not merely similar, but identical, provided that one sets the hard sphere diameter
$a_{\text{h.s.}}=x_{\text{node}}$.
We prove this first in the simplest case $N=2$. It is sufficient to restrict ourselves to the region $x\ge x_{\text{node}}$;
the solution for $x\le -x_{\text{node}}$ can then be obtained by replacing the argument $x$ by $|x|$. 
The parabolic cylinder function $D_\nu$ defined by Eq.~(\ref{parabolic}) \cite{Gra80,MorFes53,Cir01,Fra03,TemSolSch08} satisfies 
the correct Schr\"{o}dinger equation for $x>x_{\text{node}}$ as well as the correct boundary conditions of vanishing at 
$|x|=x_{\text{node}}$ and infinity and is normalizable, so it is certainly an allowable energy eigenstate for hard spheres of
diameter $x_{\text{node}}$. The only remaining question is whether it is the $\emph{ground}$ state for that given hard sphere diameter.
Since Schr\"{o}dinger's equation for the given region and boundary conditions is a well-posed Dirichlet problem and the solution
given by $D_\nu$ is nodeless in this region, it follows from Sturm-Lioville theory that it is the hard sphere
ground state. At $|\lambda|=\infty$, $D_\nu$ reduces to the TG ground state
\cite{GirWriTri01}, which is indeed the ground state for hard spheres of diameter $x_{\text{node}}=0$ zero. As $|\lambda|$ decreases, 
this node moves out from the origin, and the solution given by $D_\nu$ for
$x\ge x_{\text{node}}$ continues to be the ground state for $|\lambda|<\infty$.  
The \emph{only} difference between the sTG and HSB wave functions apart from normalization is that
the sTG wave function allows penetration into the region $|x|<a_{h.s.}$. This holds for $-\infty<\lambda<0$, 
and the penetration is small when $|\lambda|\gg 1$
so that $x_{\text{node}}=a_{\text{h.s.}}\approx a_{\text{1D}}$. Furthermore, the sTG and hard sphere solutions have the same energy.
We emphasize that when the wave function is extended into the interior region $|x|<x_{\text{node}}$ the resultant sTG wave function
is highly excited relative to the collapsed McGuire state due to the effects of the strong attraction at $x=0$, but when restricted to
the region $|x|\ge x_{\text{node}}$, it is \emph{identical} with the \emph{ground} state of the hard sphere Bose gas apart from
normalization, and the energies of the \emph{highly excited} sTG state and the hard sphere \emph{ground} state are exactly equal 
\cite{Note0}.

{\it Generalization to $N>2$:} The ground state in the $(x_1,\cdots,x_N)$ space including the interaction points $x_j=x_\ell$ 
is the trapped version of McGuire's cluster state, and one expects that
for $a_{\text{1D}}\to 0+$ it will be well approximated by
\begin{equation}
\psi_{B0}\approx\left[\prod_{1\le j<\ell\le N}\exp\left(-\frac{|x_j-x_l|}{a_{\text{1D}}}\right)\right]
\prod_{j=1}^N\exp\left(-\frac{x_j^2}{2x_{\text{osc}}^2}\right)\ .
\label{eq:McGuire}
\end{equation}
In fact, the LL contact conditions \cite{LieLin63} arising from the
delta interactions $g_B\delta(x_j-x_\ell)$
are satisfied exactly at each collision point $x_j=x_\ell$.
Furthermore, this expression becomes exact at all $(x_1,\cdots,x_N)$ in both limits $a_{\text{1D}}\to 0+$ (total collapse) and
$a_{\text{1D}}\to\infty$ (trapped ideal Bose gas), as well as when $x_{\text{osc}}\to\infty$.
In view of the ubiquity of such Bijl-Jastrow pair product wave functions as exact energy eigenstates of 1D Bose and
Fermi gases \cite{Gir60,GirWriTri01,GirNguOls04,GirWri05,GirMin06,CalSut6971}, we expect that the sTG state will also
be well approximated by such a pair product. The obvious generalization of the N=2 sTG wave function to arbitrary $N$ is
\begin{equation}\label{psiBnu}
\psi_{B\nu}(x_1,\cdots,x_N)=\left[\prod_{1\le j<\ell\le N}f_\nu(|q_{j\ell}|) \right]
\prod_{j=1}^N\exp\left(-\frac{x_j^2}{2x_{\text{osc}}^2}\right)\
\end{equation}
where
$f_\nu(|q_{j\ell}|)=D_\nu(|q_{j\ell}|)e^{q_{j\ell}^{2}/4}$ and $q_{j\ell}=(x_j-x_\ell)/x_{\text{osc}}$. It also satisfies the contact
conditions exactly, and
becomes exact at all $(x_1,\cdots,x_N)$ in both limits $a_{\text{1D}}\to 0+$ (trapped TG gas, where
$\nu=1$ and $f_\nu=|q_j-q_\ell|$) and
$a_{\text{1D}}\to\infty$ (trapped ideal Bose gas, which maps to the trapped FTG gas \cite{GirWri05,GirMin06}, and where $\nu=0$
and $f_\nu=1$).

There are accurate numerical solutions
for $N=3$ which prove the existence of a gas-like excited state with nodes \emph{only} at a nearest neighbor separation
$|x_{j\ell}|=x_{\text{node}}$ which increases with decreasing
$|\lambda|$, is very close to $a_{1D}$ for $|\lambda|\gg 1$, and goes to zero along with $a_{1D}$ in the TG limit $|\lambda|\to\infty$;
see Fig. 4 of \cite{TemSolSch08}. Furthermore, for all $N\ge 2$ the approximate wave functions (\ref{psiBnu}) vanish only at
$|x_{j\ell}|=x_{\text{node}}$ and become exact both at the collision points $x_{j\ell}=0$ and when all $|x_{j\ell}|\to\infty$,
and we expect the exact sTG solution to have the same properties. 
We expect that the theorem on the identity of the sTG and hard sphere wave functions and energies
in the region exterior to the nodes will also hold for all $N>2$ since the exact hard sphere ground state for hard sphere diameter
$x_{\text{node}}=0$ is the $|\lambda|=\infty$ TG ground state \cite{GirWriTri01} 
and the same arguments we used for $N=2$ should apply as
$|\lambda|$ is decreased. A rigorous appeal to Sturm-Liouville theory would require a proof that for the \emph{exact, unknown}
sTG eigenfunctions the pair distribution function vanishes only at $|x_j-x_\ell|=x_{\text{node}}$, and so far this is known
only for $N=2$ and $N=3$ \cite{TemSolSch08}. Therefore we state our belief of the identity of the sTG and hard sphere wave functions
and energies for all $N$ as a conjecture.

\begin{figure}
\begin{center}
\includegraphics[angle=-90,width=0.9\columnwidth]{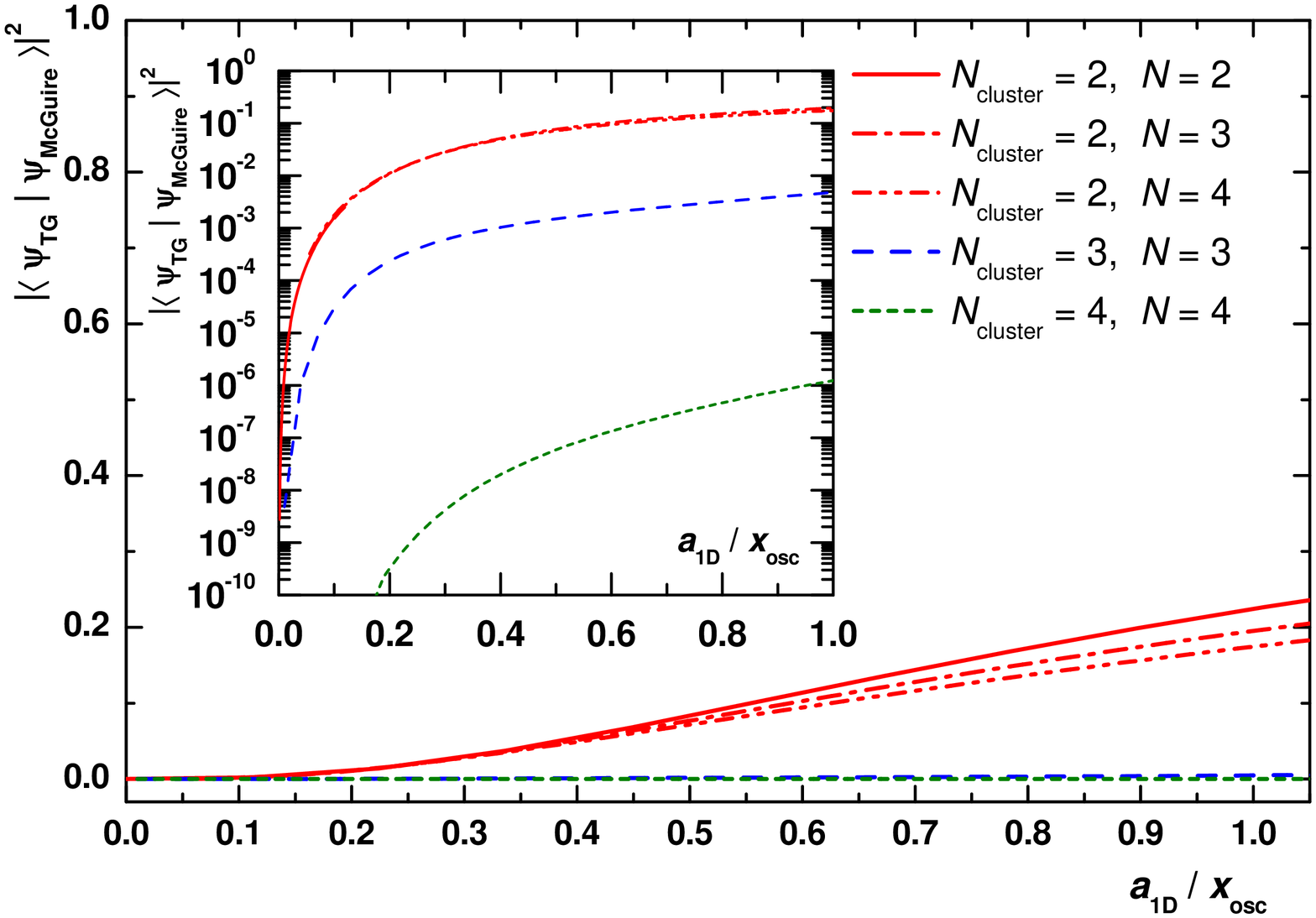}
\caption{Square of overlap integral of TG gas-like state with a McGuire state containing $N_{cl}$ clustered particles, as a function of
$a_{\text{1D}}/x_{\text{osc}}$.
Inset, the same data on a semi-logarithmic scale.}
\label{Fig1}
\end{center}
\end{figure}

{\it Sudden approximation and metastability:} In the experiment by the Innsbruck group \cite{Haletal09}, the dimensionless coupling constant $\lambda$ is changed suddenly from large positive to large negative values. For $\lambda\gg 1$ the ground state is very close to
that of the trapped TG gas, the ground state for $\lambda=+\infty$, whose \emph{exact} wave function is 
$\psi_{TG}=C_N[\prod_{1\le j<\ell\le N}|x_j-x_\ell|]\prod_{j=1}^N\exp(-x_j^2/2x_{\text{osc}}^2)$ where $C_N$ is a normalization
constant \cite{GirWriTri01}. Assuming that the switch to large negative $\lambda$ is rapid enough that the sudden approximation is valid, the
wave function after the switch will be in a superposition of all eigenstates $\psi_{\lambda\alpha}$ of the system with negative $\lambda$:
$\psi_\lambda(t)=\sum_\alpha\langle\psi_{\lambda\alpha}|\psi_{TG}\rangle\psi_{\lambda\alpha}e^{-iE_{\lambda\alpha}t/\hbar}$.
The stability of a trapped gas with attractive 1D interaction was addressed in \cite{Haletal09} by measuring the frequencies of collective breathing mode. It was found that by rapidly crossing the CIR to finite positive $a_{1D}$ the frequency increases,
which agrees with the description of the sTG state in terms of an equivalent hard rod system,
while at larger values of $a_{1D}$ the frequency decreased. These results agree with
our theorem that the sTG state has the same ground state energy as a system of hard spheres, increasing with density \cite{Note1,Note2},
so that the frequency in the sTG state increases with $a_{1D}$, although for non-zero values of $a_{1D}$ there is a possibility of populating states different from the sTG state.
We analyze the transition through the CIR in terms of the projection of the TG state to the gas-like sTG eigenstate and states that contain bound states, relying on the fact that the initial state for $\lambda\gg 1$ is very close to the $\lambda=+\infty$ TG state.
In the above expression for $\psi_\lambda(t)$ there are many-body bound states with $N=2,...,N$ particles, with the bound part described by McGuire's many-body solution Eq.~(\ref{eq:McGuire}) with a characteristic size of $a_{1D}$
The projection integral $\langle\psi_{\lambda\alpha}|\psi_{TG}\rangle$ decreases exponentially fast with the number of particles in the bound state. From this the most probable way to transfer particles to some bound state is to do so to a bound state with 2 particles.
We calculate the projection integral $\langle\psi_{\lambda\alpha}|\psi_{TG}\rangle$
for different numbers of particles and show the results in Fig.~\ref{Fig1}.
The transition to many-particle bound states is highly suppressed compared to the transition to a state where two particles are bound while all the other particles are unbound.
This can be understood my comparing characteristic volumes that a state occupies in the phase space.
The typical size of a McGuire state of $N$ particles is $a_{1D}^N$, while the TG gas occupies $(\sqrt{N}x_{osc})^N$, so that the overlap decreases dramatically as $N$ is increased.
From this the most relevant transition is to a state with two particles bound.
The overlap integral was evaluated analytically for $2$ bound particles in systems of $N=2,3,4$ particles, and numerically for
bound clusters of $3$ and $4$ particles, with 
the bound cluster described by the McGuire wave function (\ref{eq:McGuire}) while leaving the other particles in the TG
state.
In all cases the overlap is very small when $a_{\text{1D}}\ll x_{\text{osc}}$, as
in the experiment \cite{Haletal09}.

These results are expected since $\psi_{TG}$ is an exact eigenstate not only for $\lambda=+\infty$, but also for
$\lambda=-\infty$, since both map to the ideal Fermi gas \cite{Gir60,CheShi98}. Hence the system is \emph{completely stable} under a sudden jump from $\lambda=+\infty$ to $\lambda=-\infty$, and metastable following a jump from
$\lambda\gg 1$ to $\lambda\ll -1$.
\begin{acknowledgments}
This work was motivated by a talk by and discussions with Hanns-Cristoph N\"{a}gerl at the BEC2009 conference in
Sant Feliu de Guixols, Spain in September 2009, regarding the recent sTG experiment by the Innsbruck group \cite{Haletal09}.
M.D.G. also thanks Vladimir Yurovsky for a stimulating conversation there. We are very grateful to Manuel Valiente for comments
and suggestions regarding the proof of the identity of the sTG and hard sphere wave functions outside the node at $|x|=x_{\text{node}}$.
Research of M.D.G. is supported by the U.S. Army Research Laboratory and the U.S. Army Research Office under grant number W911NF-09-1-0228, and that of G.E.A. by a post doctoral fellowship by MEC (Spain), Grant No. FIS2008-04403.
\end{acknowledgments}
\end{document}